\documentstyle[12pt]{article}
\newcommand{\p}[1]{(\ref{#1})}

\newcommand{\be}{\begin{equation}}
\newcommand{\bea}{\begin{eqnarray}}
\newcommand{\ee}{\end{equation}}

\newcommand{\eea}{\end{eqnarray}}

\begin{document}
\renewcommand{\thefootnote}{\fnsymbol{footnote}}
\begin{center}
{\bf
On the Fock Space Realizations of Nonlinear Algebras
Describing the High Spin Fields in AdS Spaces.
}\vspace{0.3cm} \\

{\v{C}estm\'{\i}r Burd\'{\i}k}\footnote{E-mail: burdik@dec1.fjfi.cvut.cz}
{and Ond\v{r}ej Navr\'{a}til}\footnote{E-mail: navratil@fd.cvut.cz}
\vspace{0.3cm} \\

{\it Department of Mathematics,
Czech Technical University,\\
Trojanova 13, 120 00 Prague 2}\vspace{0.5cm} \\

A. Pashnev\footnote{E-mail: pashnev@thsun1.jinr.dubna.su}
\vspace{0.3cm} \\
{\it JINR--Bogoliubov Theoretical Laboratory,         \\
141980 Dubna, Moscow Region, Russia} \vspace{.5cm} \\

{\bf Abstract}
\end{center}
\begin{center}
{\begin{minipage}{4.2truein}
                 \footnotesize
                 \parindent=0pt

The method of construction of Fock space realizations of Lie algebras
is generalized for nonlinear algebras. We consider as an example the
nonlinear algebra of constraints which describe the
totally symmetric fields
with higher spins in the AdS space-time.

\end{minipage}}\end{center}
                 \vskip 2em \par

\renewcommand{\thefootnote}{\arabic{footnote}}
\setcounter{footnote}0\setcounter{equation}0

\section{Introduction}
It is well known that many physical systems are invariant with respect
to nonlinear symmetries. As the most known examples one can mention
the Kepler motion of the planets or the hydrogen atom. Though the general
classification of nonlinear algebras does not exists, the examples of such
algebras, as well as the description of physical systems in which they act,
one can find in the literature. Some references and rather detailed
investigation of the so called finite W-algebras see in \cite{dBHT}.

The very interesting class of nonlinear symmetries arises in the
description of the higher spin fields in the AdS space-time.
They depend on the type of the corresponding Young tableaux\cite{BMV} and
become more complicated when the number $N$ of raws is growing.
The simplest of these symmetries is connected with the Young
tableaux with $N=1$ and describes the totally symmetric
fields of higher ranks which correspond to unitary representations
in the AdS space-time of arbitrary dimension.

Starting from a given physical system with known nonlinear symmetry
one can construct the explicit expressions for the generators
of the symmetry transformations, usually in terms of the harmonic
oscillators connected with the physical observables. In the case at hand
the generators are constructed in terms of a space-time derivatives
and $N$ additional harmonic oscillators, which describe spin degrees
of freedom and can be viewed as some internal distances in the
classical models of composite particles like a discrete string\cite{GP},\cite{FI}.
No doubt that this representation of the algebra generators is not unique.
The question arises, how one can construct other representations
of nonlinear algebras? Moreover, the knowledge of such new realizations
of the algebras can help in the search of physical systems having
these algebras as a symmetries. It can also help in the investigations
of mathematical properties of these algebras.

In the present paper we propose the regular method of construction of realizations
for nonlinear algebras. We begin with the description how the method works in
the case of linear algebras, \cite{B},\cite{BPT}, and after that
generalize it for the simplest case ($N=1$) of totally symmetric fields
describing the higher spins in the AdS space-time. For this goal we
construct the Verma module for this nonlinear algebra. The representation of
the algebra generators is constructed in terms of some auxiliary harmonic oscillators
as a result of establishing the one-to-one correspondence of the Verma module vectors
with the vectors of the Fock space generated by these oscillators.
In conclusions we mention some interesting mathematical problems
connected with the construction and some possible developments.

\setcounter{equation}0\section{Construction of the Fock space realizations
in the Lie algebra case}
In this section we describe the method, based on the construction
 given in \cite{B},\cite{BPT}.
Let ${\hat H}^i,\;(i=1,...,k)$ and ${\hat E}^{\alpha}$
be the Cartan generators and root vectors
of the algebra with the following commutation relations
\begin{eqnarray}
\label{commutator}
&&\left[{\hat H}^i,{\hat E}^\alpha\right]=\alpha(i) {\hat E}^\alpha,\\
&&\left[{\hat E}^\alpha,{\hat E}^{-\alpha}\right]=\sum\alpha^i {\hat H}^i,\\
&&\left[{\hat E}^\alpha,{\hat E}^{\beta}\right]=N^{\alpha\beta}
{\hat E}^{\alpha+\beta}.
\end{eqnarray}
Roots $\alpha(i)$ and parameters $\alpha^i,\; N^{\alpha\beta}$
are structure constants of the algebra in the Cartan -- Weyl basis.

Consider the highest weight representation of the algebra under
consideration with the highest weight vector ${|0\rangle_V}$, annihilated
by the positive roots
\begin{equation}\label{V1}
{E^\alpha|0\rangle_V}=0
\end{equation}
and being the proper vector of the Cartan generators
\begin{equation}\label{V2}
H^i{|0\rangle_V}=h^i{|0\rangle_V}.
\end{equation}

The representation which is given by \p{V1} and \p{V2}
 in the mathematical literature is called  the Verma
module \cite{D}. Following the Poincare -- Birkhoff -- Witt theorem,
 the basis space of this representation
is given by vectors
\begin{equation}\label{b1}
{\left|n_1,n_2,\dots,n_r \right \rangle_V} =
(E^{-\alpha_1})^{n_1}(E^{-\alpha_2})^{n_2}\dots (E^{-\alpha_r})^{n_r}{|0 \rangle_V}
\end{equation}
where ${\alpha_1}, {\alpha_2}, \dots {\alpha_r}$ is some ordering of positive roots
 and $n_i \in N$.

Using the commutation relations of the  algebra and the formula
$$
AB^n=\sum^n_{k=0}{n \choose k }B^{n-k}[[\dots [A,B],B] \dots ]
$$
one can calculate the action of all generators on arbitrary vector of the Verma module.
 In  \cite{B} it was shown  that,  making use of the map
\begin{equation}\label{cb2}
{\left|n_1,n_2,\dots,n_r \right \rangle_V}
\longleftrightarrow \left|n_1,n_2,\dots,n_r \right\rangle
\end{equation}
where $\left|n_1,n_2,\dots,n_r \right\rangle$   are
 base vectors of the Fock space
  \begin{equation}\label{b2}
\left|n_1,n_2,\dots,n_r \right\rangle =
(b^+_1)^{n_1}(b^+_2)^{n_2}\dots (b^+_r)^{n_r}|0 \rangle.
\end{equation}
generated by creation and annihilation operators
$b_i^+, b_i \; i=1,2,\dots,r$ with the standard commutation relations
\begin{equation}\label{heisenberg}
\left[b_i,b_j^+ \right] =\delta_{ij},
\end{equation}
the generators of the algebra in the Fock space can be written as polynomials in
creation operators.

As an explicit example of the construction  given above
let us consider the representations
of the algebra, which can be used for the
description of the higher spin fields in the flat space-time.
This algebra is linear and its generators
$L_1^+, L_2^+, L_1, L_2, G_0, L_0,$
which single out the irreducible
representation of the Poincare group have the following
 commutation relations
 \begin{eqnarray} \label{al}
 &&[L^+_1, L_1] = -L_0,  \quad
 [L^+_1, L_2] = -L_1, \quad
 [L^+_2, L_1] = -L^+_1, \nonumber \\
 &&[L_0, L_1] = 0, \quad [L_0, L_2] = 0, \quad [L^+_1, L^+_2] = 0, \\
 &&[L^+_1, L_0] = 0, \quad [L_1, L_0] = 0,  \quad [L_1, L_2] = 0, \nonumber \\
 &&[G_0, L_1] = -L_1, \quad [L^+_1, G_0] = -L^+_1, \quad [L_0, G_0] = 0, \nonumber \
 \end{eqnarray}
 with the $so(2,1)$ subalgebra
 \begin{equation} \label{so21}
 [G_0, L_2] = -2L_2, \quad
 [L^+_2, G_0] = -2L^+_2, \quad
[L^+_2, L_2] = -G_0.
\end{equation}
The defining relations for the highest vector of the Verma module are:
$$L_1 {|0\rangle_V} = 0, L_2 {|0\rangle_V} = 0, G_0 {|0\rangle_V} = h {|0\rangle_V},
L_0 {|0\rangle_V} = e {|0\rangle_V}.$$

As the base of the representation space one takes the following
vectors
\begin{equation}\label{V}
{|n_1, n_2 \rangle_V}=
(L_1^+)^{n_1}(L_2^+)^{n_2}{|0\rangle_V}
\end{equation}
and after the simple calculations  one obtains
\begin{eqnarray}\label{sv}
L_1^+\bigl|n_1,n_2\rangle_V&=&
\bigl|n_1+1,n_2\rangle_V\nonumber \\
L_2^+\bigl|n_1,n_2\rangle_V&=&
\bigl|n_1,n_2+1\rangle_V\nonumber \\
G_0\bigl|n_1,n_2\rangle_V&=&
\bigl(n_1+2n_2+h\bigr)\bigl|n_1,n_2\rangle_V\nonumber \\
L_0 \bigl|n_1,n_2\rangle_V&=&
e\bigl|n_1,n_2\rangle_V  \\
L_1\bigl|n_1,n_2\rangle_V&=&
n_2\bigl|n_1+1,n_2-1\rangle_V+en_1\bigl|n_1-1,n_2\rangle_V\nonumber \\
L_2\bigl|n_1,n_2\rangle_V&=&
n_2\bigl(h+n_1+n_2-1\bigr)\bigl|n_1,n_2-1\rangle_V+
e\frac{n_1(n_1-1)}2\bigl|n_1-2,n_2\rangle_V.\nonumber \
\end{eqnarray}

Consider the auxiliary Fock space generated by the
annihilation and creation operators $b_1,b_2,b^+_1,b^+_2$.
 If one takes the vectors in the Fock space
\begin{equation}\label{F}
{|n_1, n_2 \rangle}=
(b_1^+)^{n_1}(b_2^+)^{n_2}{|0\rangle}
\end{equation}
in one-to-one correspondence with the vectors \p{V} in the Verma module,
it can be seen that the action of the  following operators  in the
Fock space
\begin{eqnarray}\label{R1}
L_1^+&=&b_1^+\nonumber \\
L_2^+&=&b_2^+\nonumber \\
G_0&=&b_1^+b_1+2b_2^+b_2+h \\
L_0&=&e, \nonumber \\
L_1&=&b_1^+b_2+eb_1,\nonumber \\
L_2&=&
\bigl(b_1^+b_1+b_2^+b_2+h\bigr)b_2\nonumber. \
\end{eqnarray}
is identical to the expressions \p{sv} for the Verma module. So, \p{R1}
gives realization of the algebra in the Fock space.

\setcounter{equation}0\section{Construction of the Fock space realization
in nonlinear algebra case}

In the AdS space-time the algebra \p{al} is modified by terms proportional
to the parameter $r$ which is the square of the inverse radius of
this space. When $r\neq 0$ the algebra is nonlinear and have the following
nonzero commutation relations \cite{BMV}
\begin{eqnarray}
&&[L^+_1, L_1] = - L_0,  \quad
[L^+_1, L_2] = -L_1, \quad
[L^+_2, L_1] = -L^+_1, \nonumber \\
&&[ L_0, L_1] = -2rL_1 + 4rG_0  L_1 -
      8rL^+_1  L_2, \nonumber\\
&&[L^+_1,  L_0] = -2rL^+_1 + 4rL^+_1  G_0 -
      8rL^+_2  L_1,  \nonumber \\
&&[G_0, L_1] = -L_1, \quad [L^+_1, G_0] = -L^+_1,\label{Alg}
\end{eqnarray}
with the $so(2,1)$ subalgebra \p{so21}.
The structure of this nonlinear algebra is simplified if one introduces
new generator $C$ instead of the generator $L_0$: $L_0=C+2K$, where
$K$ is the Casimir operator of the subalgebra $so(2,1)$
\begin{equation} \label{ln}
K= 4L^+_2  L_2 - G_0
G_0 + 2G_0,
\end{equation}
The generator $C$ commute with all other generators and is the central
charge of the algebra.

It means that we are still able to define the Verma modules as
the highest weight representation of this algebra under
consideration with the highest weight vector ${|0\rangle_V}$, annihilated
by $L_1, L_2$
\begin{equation}\label{h1}
{L_1|0\rangle_V}=0, \quad {L_2|0\rangle_V}=0,
\end{equation}
and being the proper vector of the generators $G_0, C $ and,
correspondingly, of $G_0, {L}_0 $
\begin{equation}\label{h2}
G_0{|0\rangle_V}=h{|0\rangle_V}, \quad {L}_0{|0\rangle_V}=e{|0\rangle_V}.
\end{equation}

The basis of the representation space is given by vectors
$$
\bigl(L_1^+\bigr)^{n_1}\bigl(L_2^+\bigr)^{n_2}|0\rangle_V=
\bigl|n_1,n_2\rangle_V\,.
$$
For further calculations we will define the operators $K_1, K_2$ with the
help of the following relations
\begin{eqnarray}
&&\bigl[K,L_1^+\bigr]=L_1^+-2L_1^+G_0+4L_2^+L_1=K_1\nonumber \\
&&\bigl[L_0,L_1^+\bigr]=2rK_1\nonumber \\
&&\bigl[K_1,L_1^+\bigr]=4L_2^+L_0-2\bigl(L_1^+\bigr)^2=K_2\nonumber \\
&&\bigl[K_2,L_1^+\bigr]=8rL_2^+K_1\nonumber \\
&&\bigl[K_1,L_2^+\bigr]=0, \quad \bigl[K_2,L_2^+\bigr]=0
\end{eqnarray}
It is easy to see that
\begin{eqnarray}
&&K_1\bigl|0\bigr>=\bigl(1-2h\bigr)\bigl|1,0\bigr> \nonumber \\
&&K_2\bigl|0\bigr>=4e\bigl|0,1\bigr>-2\bigl|2,0\bigr> \nonumber \,.
\end{eqnarray}
To obtain the explicit form of the representation we need the commutation rules
with $(L^+_1)^n, L^+_2)^n$
\begin{eqnarray}
G_0\bigl(L_1^+\bigr)^n&=&
\bigl(L_1^+\bigr)^nG_0+n\bigl(L_1^+\bigr)^n\nonumber \\
G_0\bigl(L_2^+\bigr)^n&=&
\bigl(L_2^+\bigr)^nG_0+2n\bigl(L_2^+\bigr)^n\nonumber \\
L_0\bigl(L_1^+\bigr)^n&=&\bigr(L_1^+\bigr)^nL_0+
\sum_{k=1}\frac{(8r)^k}4{{n}\choose{2k-1}}
\bigl(L_1^+\bigr)^{n-2k+1}\bigl(L_2^+\bigr)^{k-1}K_1+\nonumber \\
&+&
\sum_{k=1}\frac{(8r)^k}4{{n}\choose{2k}}
\bigl(L_1^+\bigr)^{n-2k}\bigl(L_2^+\bigr)^{k-1}K_2\nonumber \\
L_1\bigl(L_1^+\bigr)^n&=&
\bigl(L_1^+\bigr)^nL_1+n\bigl(L_1^+\bigr)^{n-1}L_0+
\sum_{k=1}\frac{(8r)^k}4{{n}\choose{2k}}
\bigl(L_1^+\bigr)^{n-2k}\bigl(L_2^+\bigr)^{k-1}K_1+\nonumber \\
&+&
\sum_{k=1}\frac{(8r)^k}4{{n}\choose{2k+1}}
\bigl(L_1^+\bigr)^{n-2k-1}\bigl(L_2^+\bigr)^{k-1}K_2\nonumber \\
L_1\bigl(L_2^+\bigr)^n&=&
\bigl(L_2^+\bigr)^nL_1+nL_1^+\bigl(L_2^+\bigr)^{n-1}\nonumber \\
L_2\bigl(L_1^+\bigr)^n&=&
\bigl(L_1^+\bigr)^nL_2+n\bigl(L_1^+\bigr)^{n-1}L_1+
{{n}\choose 2}\bigl(L_1^+\bigr)^{n-2}L_0+\nonumber \\
&+&
\sum_{k=1}\frac{(8r)^k}4{{n}\choose{2k+1}}
\bigl(L_1^+\bigr)^{n-2k-1}\bigl(L_2^+\bigr)^{k-1}K_1+\nonumber \\
&+&
\sum_{k=1}\frac{(8r)^k}4{{n}\choose{2k+2}}
\bigl(L_1^+\bigr)^{n-2k-2}\bigl(L_2^+\bigr)^{k-1}K_2\nonumber \\
L_2\bigl(L_2^+\bigr)^n&=&
\bigl(L_2^+\bigr)^nL_2+n\bigl(L_2^+\bigr)^{n-1}G_0+
n(n-1)\bigl(L_2^+\bigr)^{n-1}
\end{eqnarray}

With the help of these formulas we obtain explicitly the Verma module representation
\begin{eqnarray}
L_1^+\bigl|n_1,n_2\rangle_V&=&
\bigl|n_1+1,n_2\rangle_V\nonumber \\
L_2^+\bigl|n_1,n_2\rangle_V&=&
\bigl|n_1,n_2+1\rangle_V\nonumber \\
G_0\bigl|n_1,n_2\rangle_V&=&
\bigl(n_1+2n_2+h\bigr)\bigl|n_1,n_2\rangle_V\nonumber \\
L_0\bigl|n_1,n_2\rangle_V&=&
e\bigl|n_1,n_2\rangle_V+
(1-2h)\sum_{k=1}\frac{(8r)^k}4{{n_1}\choose{2k-1}}
\bigl|n_1-2k+2,n_2+k-1\rangle_V+\nonumber \\
&+&
\sum_{k=1}\frac{(8r)^k}4{{n_1}\choose{2k}}
\Bigl(4e\bigl|n_1-2k,n_2+k\rangle_V-
2\bigl|n_1-2k+2,n_2+k-1\rangle_V\Bigr)\nonumber \\
L_1\bigl|n_1,n_2\rangle_V&=&
n_2\bigl|n_1+1,n_2-1\rangle_V+en_1\bigl|n_1-1,n_2\rangle_V+\nonumber \\
&+&
\frac{(1-2h)}4\sum_{k=1}(8r)^k{{n_1}\choose{2k}}
\bigl|n_1-2k+1,n_2+k-1\rangle_V+\nonumber \\
&+&
e\sum_{k=1}(8r)^k{{n_1}\choose{2k+1}}
\bigl|n_1-2k-1,n_2+k\rangle_V-\nonumber \\
&-&
\frac12\sum_{k=1}(8r)^k{{n_1}\choose{2k+1}}
\bigl|n_1-2k+1,n_2+k-1\rangle_V\nonumber \\
L_2\bigl|n_1,n_2\rangle_V&=&
n_2\bigl(h+n_1+n_2-1\bigr)\bigl|n_1,n_2-1\rangle_V+
e\,\frac{n_1(n_1-1)}2\bigl|n_1-2,n_2\rangle_V+\nonumber \\
&+&
\frac{(1-2h)}4\sum_{k=1}(8r)^k{{n_1}\choose{2k+1}}
\bigl|n_1-2k,n_2+k-1\rangle_V+\nonumber \\
&+&
e\sum_{k=1}(8r)^k{{n_1}\choose{2k+2}}
\bigl|n_1-2k-2,n_2+k\rangle_V-\nonumber \\
&-&
\frac12\sum_{k=1}(8r)^k{{n_1}\choose{2k+2}}
\bigl|n_1-2k,n_2+k-1\rangle_V
\end{eqnarray}
Now again if we use the operators $b_1,b_2,b^+_1,b^+_2$ we obtain
the Fock space realization of the algebra \p{Alg}
\begin{eqnarray}
L_1^+&=&b_1^+\nonumber \\
L_2^+&=&b_2^+\nonumber \\
G_0&=&b_1^+b_1+2b_2^+b_2+h \\
L_0&=&\sum_{k=0}(8r)^k
\left(\frac{e}{(2k)!}-\frac{2r(2h-1)}{(2k+1)!}b_1^+b_1-
\frac{4r}{(2k+2)!}\bigl(b_1^+\bigr)^2b_1^2\right)
b_1^{2k}\bigl(b_2^+\bigr)^k\nonumber \\
L_1&=&b_1^+b_2+
\sum_{k=0}(8r)^k\left(\frac{e}{(2k+1)!}-
\frac{2r(2h-1)}{(2k+2)!}b_1^+b_1-
\frac{4r}{(2k+3)!}\bigl(b_1^+\bigr)^2b_1^2\right)
b_1^{2k+1}\bigl(b_2^+\bigr)^k\nonumber \\
L_2&=&
\bigl(b_1^+b_1+b_2^+b_2+h\bigr)b_2+\nonumber \\
&+&
\sum_{k=0}(8r)^k\left(\frac{e}{(2k+2)!}-
\frac{2r(2h-1)}{(2k+3)!}b_1^+b_1-
\frac{4r}{(2k+4)!}\bigl(b_1^+\bigr)^2b_1^2\right)
b_1^{2k+2}\bigl(b_2^+\bigr)^k \nonumber \
\end{eqnarray}
The infinite sums in this expressions are rather simple. For example, first
of them can be written in terms of the formal variable $x=\sqrt{8rb_1^2b_2}$
as
$$
e\sum_{k=0}
\frac{(8r)^k}{(2k)!}
b_1^{2k}\bigl(b_2^+\bigr)^k= e \cosh x.
$$
All the other sums can be also written in terms of the $\cosh x$ and $\sinh x$.

\section{Conclusions}
In this paper we have generalized the method of constructions of Fock space
representations for nonlinear algebras, developed for linear algebras in
\cite{B},\cite{BPT}. The method uses the notion of the Verma module for
an algebra under consideration. Taking as an example the very important
physically case of higher spin fields in the AdS space-time we
constructed the Verma module, as well as the Fock space representation
for the generators of the corresponding nonlinear algebra.

From the mathematical point of view the construction of the Verma module
gives the possibility to analyze its structure and search for the
singular vectors in it. The general procedure will then help to construct
the finite dimensional representations of considered nonlinear algebra.
It would be interesting to carry out investigations along this line,
as well as make the generalization on the case of more complicated
nonlinear algebras connected with irreducible unitary representations
in the AdS space-time of arbitrary dimensions, described by the Young
tableaux with more then one raw.

\noindent {\bf Acknowledgments.}
The work of A.P. was supported in part by
the INTAS grant, project No 00-00254 and by the RFBR and joint DFG-RFBR grants.
The research was also partially supported by Grant GACR 201/01/0130
and Votruba-Blokhintcev (Czech Republic-BLTP JINR) grant.

\end{document}